\documentclass[12pt]{article}
\usepackage{amsmath,amssymb,amsfonts,color,graphicx,cite,color,feynarts}
\usepackage{latexsym}
\usepackage[normalem]{ulem}
\usepackage{epsfig,psfrag,rotating,soul}
\usepackage{rotfloat}
\usepackage{orcidlink}
\input paperdef

\evensidemargin \oddsidemargin
\allowdisplaybreaks

\begin{document}
\thispagestyle{empty}

\def\thefootnote{\fnsymbol{footnote}}


\vspace{0.5cm}

\begin{center}

\begin{large}
\textbf{Higgs Decay to $Z\gamma$ in the Minimal Supersymmetric Standard}
\\[2ex]
\textbf{Model and Its Nonholomorphic Extension}
\end{large}

\vspace{1cm}
{\sc
S.~Israr$^{1}$%
\footnote{email: sajidisrar7@gmail.com}%
and  M.~Rehman\orcidlink{0000-0002-1069-0637}$^{1}$%
\footnote{email: m.rehman@comsats.edu.pk}%

}

\vspace*{.7cm}
{\sl
${}^1$Department of Physics, Comsats University Islamabad, 44000
  Islamabad, Pakistan \\[.1em] 
}
\end{center}

\vspace*{0.1cm}

\begin{abstract}
\noindent

Recent measurements of \( h \rightarrow Z \gamma \) from CMS and ATLAS indicate an excess over Standard Model (SM) predictions, suggesting the presence of new physics. In this work, we investigate the \( h \rightarrow Z \gamma \) decay within the minimal supersymmetric standard model (MSSM) and its non-holomorphic extension, the NHSSM. Calculations are performed using the \fa/\fc ~setup, utilizing the pre-existing model file for MSSM and generating the NHSSM model file with the support of {\tt SARAH}. In the allowed parameter space, MSSM contributions to \( h \rightarrow Z \gamma \) can significantly surpass SM predictions reaching to the value $\Gamma(h \rightarrow Z \gamma)= 9.77 \times 10^{-6} \gev$, thereby bringing \( \Gamma (h \rightarrow Z \gamma) \) closer to the experimental value. The SM predicted value deviates from the experimental value by $1.7 \sigma$. However, the MSSM contributions can reduce this deviation to less than $1 \sigma$. In contrast, NHSSM contributions remain negligible and do not produce sizable corrections to the \( h \rightarrow Z \gamma \) decay width.

\end{abstract}

\def\thefootnote{\arabic{footnote}}
\setcounter{page}{0}
\setcounter{footnote}{0}

\newpage



\section{Introduction}
\label{sec:intro}

The Minimal Supersymmetric Standard Model (MSSM) \cite{Fayet:1974pd,Fayet:1976et,Fayet:1977yc, Nilles:1983ge, Haber:1984rc,Barbieri:1987xf} stands out as one of the most compelling and extensively researched extensions of the Standard Model (SM). By introducing new particles, it addresses several unresolved issues within the SM framework. However, the conventional MSSM framework faces increasing constraints from data provided by the Large Hadron Collider (LHC), prompting researchers to explore alternative or modified models.

One promising avenue involves incorporating nonholomorphic soft supersymmetry breaking (SSB) terms \cite{Girardello:1981wz, Bagger:1993ji}. This modification gives rise to a variant known as the nonholomorphic supersymmetric standard model (NHSSM) \cite{Cakir:2005hd}. The NHSSM presents intriguing phenomenological implications that may reconcile supersymmetric (SUSY) predictions with experimental observations. This approach not only expands the theoretical possibilities within supersymmetry but also offers potential solutions to discrepancies observed between MSSM predictions and experimental results\cite{Rehman:2022ydc,Jack:1999ud,Jack:1999fa,Cakir:2005hd,Un:2014afa,Chattopadhyay:2016ivr, Chattopadhyay:2017qvh,Un:2023wws}

Recently, the ATLAS and CMS collaborations have announced evidence for the \( h \rightarrow Z \gamma \) decay with a statistical significance of \( 3.4 \sigma \) \cite{ATLAS:2023yqk}. The measured signal yield is \( \mu = 2.2 \pm 0.7 \) times the SM prediction\cite{Djouadi:1997yw,LHCHiggsCrossSectionWorkingGroup:2016ypw}. The measured branching fraction is ${\rm Br}(h \rightarrow Z \gamma)_{\rm Exp}= (3.4 \pm 1.1) \times 10^{-3}$ more than twice larger than the SM prediction given by ${\rm Br}(h \rightarrow Z \gamma)_{\rm SM}= (1.5 \pm 0.1) \times 10^{-3}$. This decay channel has been calculated very precisely in the SM \cite{Cahn:1978nz, Bergstrom:1985hp, Spira:1991tj, Bonciani:2015eua, Gehrmann:2015dua,Buccioni:2023qnt, Sang:2024vqk}, making this deviation a potential hint for physics beyond the SM. This is a loop-induced decay, which does not occur at tree level. Consequently, new physics contributions to this decay could match or even exceed the SM predictions.

The \( h \rightarrow Z \gamma \) decay has been extensively studied in various new physics models \cite{Hue:2017cph, Chiang:2012qz, Maru:2013bja, Yue:2013qba, Fontes:2014xva, Hung:2019jue,Boto:2023bpg}. In supersymmetric scenarios, detailed investigations have been conducted \cite{Weiler:1988xn, Cao:2013ur, Belanger:2014roa, Hammad:2015eca, Liu:2020nsm}. Specifically, in \cite{Cao:2013ur}, it was proposed that contributions from the general MSSM could result in deviations of up to 10\% from the SM predictions. 
To the best of our knowledge, the \( h \rightarrow Z \gamma \) decay has not yet been evaluated within the framework of the NHSSM. This represents an unexplored area that could potentially offer new insights into the behavior of the Higgs boson in alternative supersymmetric models. Investigating this decay within the NHSSM could provide valuable information on how nonholomorphic terms affect the predictions and whether they can account for the observed deviations from the SM.

In this paper, we have analysed the MSSM contributions to the \( h \rightarrow Z \gamma \) decay at the one-loop level and found significant differences compared to earlier studies \cite{Cao:2013ur}. Additionally, we have calculated the \( h \rightarrow Z \gamma \) decay width within the framework of the NHSSM. 
Our analysis was conducted using the \fa/\fc ~setup~\cite{Hahn:2000kx,Hahn:2001rv,Fritzsche:2013fta,Hahn:1998yk}, utilizing the preexisting model file for the MSSM and developing a new \fa ~model file for the NHSSM. The NHSSM model file was developed with the help of the Mathematica package {\tt SARAH} \cite{Staub:2009bi, Staub:2010jh, Staub:2012pb, Staub:2013tta, Staub:2015kfa}. This comprehensive approach allowed us to explore the impact of nonholomorphic terms on the \( h \rightarrow Z \gamma \) decay and to compare the results with those from the MSSM.

The structure of the paper is as follows: we begin by outlining the key characteristics of the NHSSM in \refse{sec:model_NHSSM}. Next, we detail the computational framework in \refse{sec:CalcSetup}, followed by the presentation of numerical results in \refse{sec:NResults}. Finally, we summarize our findings and draw conclusions in \refse{sec:conclusions}.

\section{Model set-up}
\label{sec:model_NHSSM}

The MSSM represents the most straightforward and fundamental supersymmetric extension of the SM particle content. It serves as the foundational framework for incorporating supersymmetry into the existing SM, ensuring minimal complexity while achieving supersymmetric objectives. 

The MSSM features an extended Higgs boson sector, comprising five distinct particles. These include two $\cp$-even Higgs bosons denoted as $h$ and $H$, with $h$ corresponding to the Higgs boson discovered at the LHC. Additionally, there is a $\cp$-odd Higgs boson $A$ and a pair of charged Higgs bosons $H^{\pm}$. In its scalar sector, the MSSM incorporates scalar quarks and scalar leptons as well. Furthermore, in the fermionic sector, the MSSM introduces additional particles known as neutralinos and charginos.

If supersymmetry were an exact symmetry of nature, it would necessitate that supersymmetric particles have the same masses as their SM counterparts. Under these conditions, we would have already detected supersymmetric particles at collider experiments. Despite numerous efforts, no supersymmetric particle has been discovered, indicating that supersymmetry likely exists as a broken symmetry. To effectively break supersymmetry and achieve realistic low-energy phenomenology, the MSSM relies on a series of soft SUSY-breaking (SSB) parameters. These parameters include mass terms, trilinear couplings, and bilinear terms, among others. The general setup for these SSB parameters is given as follows:
\begin{eqnarray}
\label{softbreaking}
-\cL_{\rm soft}&=&(m_{\tilde Q}^2)_i^j {\tilde q}_{L}^{\dagger i}
{\tilde q}_{Lj}
+(m_{\tilde u}^2)^i_j {\tilde u}_{Ri}^* {\tilde u}_{R}^j
+(m_{\tilde d}^2)^i_j {\tilde d}_{Ri}^* {\tilde d}_{R}^j
\nonumber \\
& &+(m_{\tilde L}^2)_i^j {\tilde l}_{L}^{\dagger i}{\tilde l}_{Lj}
+(m_{\tilde e}^2)^i_j {\tilde e}_{Ri}^* {\tilde e}_{R}^j
\nonumber \\
& &+{\tilde m}^2_{1}h_1^{\dagger} h_1
+{\tilde m}^2_{2}h_2^{\dagger} h_2
+(B \mu h_1 h_2
+ {\rm h.c.})
\nonumber \\
& &+ ( A_d^{ij}h_1 {\tilde d}_{Ri}^*{\tilde q}_{Lj}
+A_u^{ij}h_2 {\tilde u}_{Ri}^*{\tilde q}_{Lj}
+A_l^{ij}h_1 {\tilde e}_{Ri}^*{\tilde l}_{Lj}
\nonumber \\
& & +\frac{1}{2}M_1 {\tilde B} {\tilde B}
+\frac{1}{2}M_2 {\tilde W} {\tilde W}
+\frac{1}{2}M_3 {\tilde G} {\tilde G} + {\rm h.c.}).
\end{eqnarray}
Here, \( m_{\tilde{Q}}^2 \) and \( m_{\tilde{L}}^2 \) are \( 3 \times 3 \) matrices in family space, with \( i \) and \( j \) representing the generation indices. These matrices correspond to the SSB masses for the left-handed squark \( \tilde{q}_{L} \) and slepton \( \tilde{l}_{L} \) SU(2) doublets, respectively. The terms \( m_{\tilde{u}}^2 \), \( m_{\tilde{d}}^2 \), and \( m_{\tilde{e}}^2 \) represent the soft masses for the right-handed up-type squark \( \tilde{u}_{R} \), down-type squark \( \tilde{d}_{R} \), and charged slepton \( \tilde{e}_{R} \) SU(2) singlets, respectively. The matrices \( A_u \), \( A_d \), and \( A_l \) are \( 3 \times 3 \) matrices that describe the trilinear couplings for up-type squarks, down-type squarks, and charged sleptons, respectively. The parameter \( \mu \) denotes the Higgs mixing parameter, while \( \tilde{m}_1 \), \( \tilde{m}_2 \), and \( B \) are the SSB parameters associated with the Higgs sector, where \( h_1 \) and \( h_2 \) indicate the two Higgs doublets. Finally, \( M_1 \), \( M_2 \), and \( M_3 \) define the mass terms for the bino, wino, and gluino, respectively.

In the MSSM, the superpotential must be holomorphic, leading to the SSB sector typically being parameterized using holomorphic operators. However, the MSSM can be extended by incorporating nonholomorphic terms in the SSB sector.~\cite{Girardello:1981wz, Bagger:1993ji, Chakrabortty:2011zz}. 
In its simplest form the following terms can be introduced in
the SSB sector of the MSSM: 
\begin{eqnarray}
\label{NonH-TrilinearTerms}
-\cL_{\rm soft}^{\rm NH}&=&A_d^{^\prime ij}h_2 {\tilde d}_{Ri}^*{\tilde q}_{Lj}
+A_u^{^\prime ij}h_1 {\tilde u}_{Ri}^*{\tilde q}_{Lj}
+A_l^{^\prime ij}h_2 {\tilde e}_{Ri}^*{\tilde l}_{Lj}
+\mu^{\prime} {\tilde h}_1 {\tilde h}_2
\end{eqnarray}
Here $\mu^{\prime}$ is nonholomorphic higgsino mass term, whereas $A_u^{^\prime}$,
$A_d^{^\prime}$ and $A_l^{^\prime}$ denote the nonholomorphic trilinear
coupling matrices for up-type squarks, down-type squarks, and charged
sleptons, respectively. These terms do not necessarily have any
relationship with the holomorphic trilinear soft terms given in
\refeq{softbreaking}.

With the inclusion of these nonholomorphic trilinear terms, the sfermion mass matrix can be expressed as,  
\begin{equation}
M_{\tilde{f}}^{2}=\left(
\begin{array}
[c]{cc}%
m_{\tilde{f}LL}^{2} & m_{\tilde{f}LR}^{2}\\[.5em]
m_{\tilde{f}LR}^{2\dag} & m_{\tilde{f}^{\prime}RR}^{2}%
\end{array}
\right) \label{fermion mass matrix}%
\end{equation}
with%
\begin{align}
m_{\tilde{f}LL}^{2}  & =m_{\tilde{f}}^{2}+M_{Z}^{2}\cos2\beta\left(  I_{3}%
^{f}-Q_{f}s_{W}^{2}\right)  +m_{f}^{2}\nonumber\\
m_{\tilde{f}^{\prime}RR}^{2}  & =m_{\tilde{f}^{%
\acute{}%
}}^{2}+M_{Z}^{2}\cos2\beta Q_{f^{\prime}}s_{W}^{2}+m_{f}^{2}\nonumber\\
m_{\tilde{f}LR}^{2}  & =m_{f}X_{f}\text{ ;
\ \ \ }X_{f}=A_{f}-(\mu+A_f^\prime)\left\{  \cot\beta;\tan\beta\right\} \label{mass terms}%
\end{align}

The parameter $\cot{\beta}$ corresponds to the up-type and $\tb$ to the down-type sfermions. The $I_{3}^{f}$ represents the weak isospin of fermions, $Q_{f}$ stands for the electromagnetic charge, and $m_{f}$ denotes the mass of SM fermions. $\MZ$ and $\MW$ correspond to the masses of the $Z$ and $W$ bosons, respectively, while $s_W$ is defined as the square root of $1 - c_W^2$, with $c_W = \MW/\MZ$. The parameter $\tb$ is defined as the ratio of the vacuum expectation values of the two Higgs doublets, denoted by $v_1$ and $v_2$, such that $\tb = v_2 / v_1$.

\section{Computation of $h \rightarrow Z \gamma$}
\label{sec:CalcSetup}

\subsection{Experimental status of $h \rightarrow Z \gamma$}
\label{Exp-status}
The $h \rightarrow Z\gamma $ decay stands out from other Higgs decays observed so far because its final state does not consist of two identical particles (such as \(\gamma\gamma\)) or a particle-antiparticle pair (such as \(b\overline{b}\)). Furthermore, since it is a loop-induced process, the $h \rightarrow Z\gamma $ decay serves as an important probe of the SM at the quantum level, providing insights into the Higgs boson's properties and its interactions with heavy particles\cite{Djouadi:2005gi}.

The ATLAS and CMS collaborations have recently provided evidence for the $h \rightarrow Z\gamma $ decay, achieving a statistical significance of \( 3.4 \sigma \) \cite{ATLAS:2023yqk}.  The measured branching fraction is 
\begin{equation}
{\rm Br}(h \rightarrow Z \gamma)_{\text{Exp}} = (3.4 \pm 1.1) \times 10^{-3}.    
\end{equation}
Currently, the measurement of the Higgs boson's full width has a substantial uncertainty\cite{ParticleDataGroup:2022pth}, given by:

\begin{equation}
\Gamma_h = 3.2^{+2.4}_{-1.7} \text{ MeV}
\end{equation}
To avoid this significant uncertainty, we use the more accurate theoretical prediction for the full width \cite{LHCHiggsCrossSectionWorkingGroup:2016ypw}:

\begin{equation}
\Gamma_h = 4.07^{+4.0\%}_{-3.9\%} \text{ MeV}
\label{HfullWidth-SM}
\end{equation}
The SM prediction for the branching ratio of the Higgs boson decaying to a \(Z\) boson and a photon is:

\begin{equation}
\text{Br}(h \rightarrow Z \gamma)_{\text{SM}} = (1.5 \pm 0.1) \times 10^{-3}
\end{equation}

The experimentally observed value exceeds the SM prediction by more than a factor of two. In terms of decay width, using \refeq{HfullWidth-SM}, the SM prediction can be expressed as:
\begin{equation}
\Gamma(h \rightarrow Z \gamma)_{\text{SM}} = (6.1 \pm 0.4) \times 10^{-6} \text{ GeV}
\end{equation}
We define the relative deviation between the SM prediction and the experimentally measured value using the following equation:
\begin{equation}
{\rm Dev}(Z\gamma)_{\rm SM} = \frac{\Gamma(h \rightarrow Z \gamma)_{\text{SM}}}{\Gamma(h \rightarrow Z \gamma)_{\text{Exp}}} - 1
\end{equation}
This results in a relative deviation of approximately ${\rm Dev}(Z\gamma)_{\rm SM} \approx -56\%$. Similarly, we define the relative deviation of the MSSM prediction compared to the experimental value as:
\begin{equation}
{\rm Dev}(Z\gamma)_{\rm MSSM} = \frac{\Gamma(h \rightarrow Z \gamma)_{\text{MSSM}}}{\Gamma(h \rightarrow Z \gamma)_{\text{Exp}}} - 1
\label{Eq:Dev-MSSM}
\end{equation}

\subsection{Computational Setup}

Computer programs \fa\ and \fc~\cite{Hahn:2000kx,Hahn:2001rv,Fritzsche:2013fta,Hahn:1998yk} are designed to assist in calculating analytical and numerical results for decay processes and scattering cross-sections. These packages are especially beneficial when working within the frameworks of the SM or MSSM, as \fa\ includes built-in model files for these theories. One-loop numerical evaluations can be conducted using the \fa/\fc~ framework, which interfaces with the computer package {\tt LoopTools}\cite{Hahn:1998yk}. This package provides the essential formulas for one-loop integrals required for the calculations.

\begin{figure}[htb!]
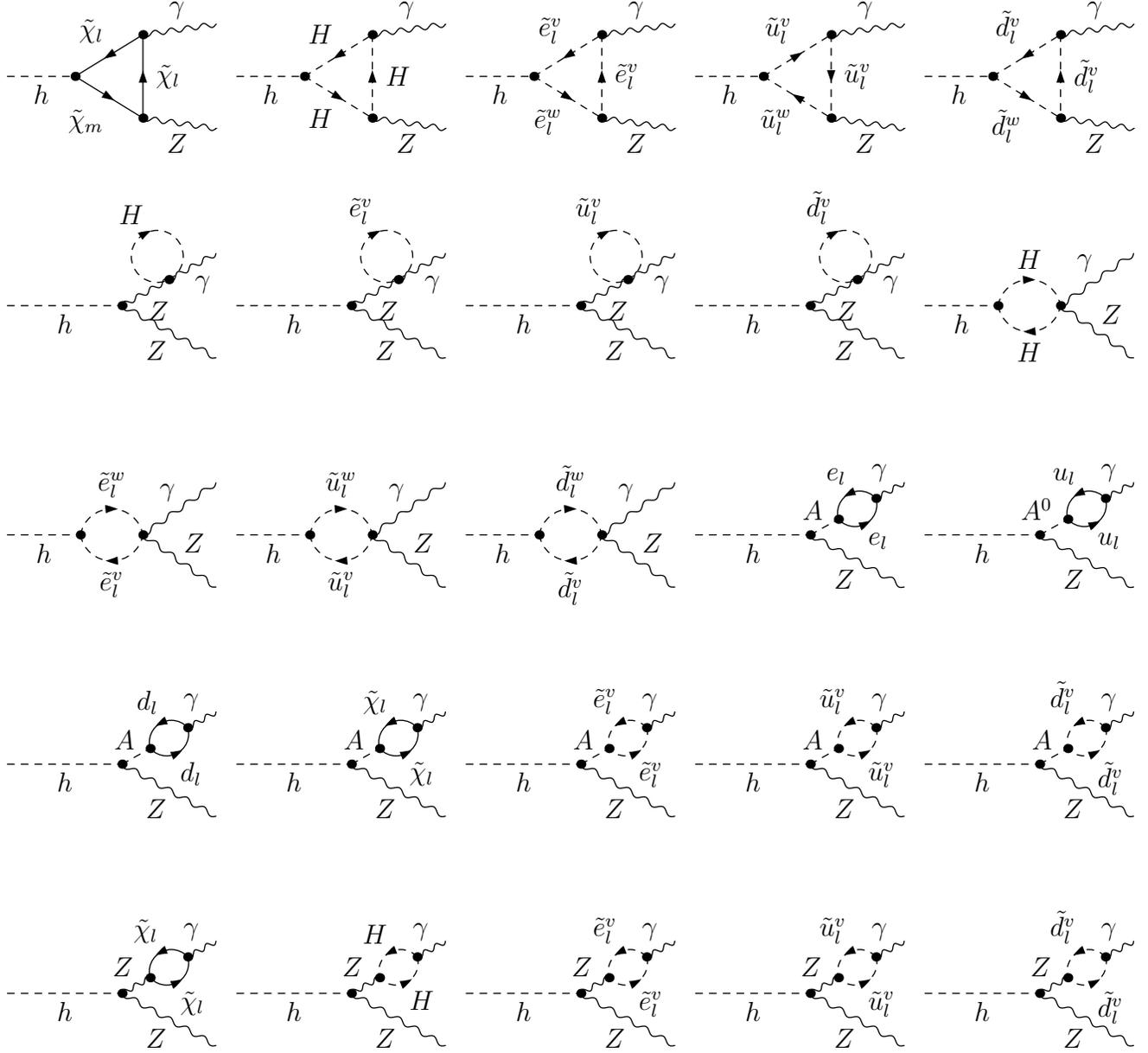

\begin{center}

\unitlength=1.0bp%
\begin{feynartspicture}(500,650)(5,5)

\FADiagram{}
\FAProp(0.,10.)(6.5,10.)(0.,){/ScalarDash}{0}
\FALabel(3.25,9.18)[t]{$h$}
\FAProp(20,15.)(13.,14.)(0.,){/Sine}{0}
\FALabel(16.2808,15.5544)[b]{$\gamma$}
\FAProp(20,5.)(13.,6.)(0.,){/Sine}{0}
\FALabel(16.2808,4.44558)[t]{$Z$}
\FAProp(6.5,10.)(13.,14.)(0.,){/Straight}{-1}
\FALabel(9.2078,13.1811)[br]{$\tilde \chi_l$}
\FAProp(6.5,10.)(13.,6.)(0.,){/Straight}{1}
\FALabel(9.2078,6.81893)[tr]{$\tilde \chi_m$}
\FAProp(13.,14.)(13.,6.)(0.,){/Straight}{-1}
\FALabel(14.2741,10.)[l]{$\tilde \chi_l$}
\FAVert(6.5,10.){0}
\FAVert(13.,14.){0}
\FAVert(13.,6.){0}

\FADiagram{}
\FAProp(0.,10.)(6.5,10.)(0.,){/ScalarDash}{0}
\FALabel(3.25,9.18)[t]{$h$}
\FAProp(20,15.)(13.,14.)(0.,){/Sine}{0}
\FALabel(16.2808,15.5544)[b]{$\gamma$}
\FAProp(20,5.)(13.,6.)(0.,){/Sine}{0}
\FALabel(16.2808,4.44558)[t]{$Z$}
\FAProp(6.5,10.)(13.,14.)(0.,){/ScalarDash}{-1}
\FALabel(9.2078,13.1811)[br]{$H$}
\FAProp(6.5,10.)(13.,6.)(0.,){/ScalarDash}{1}
\FALabel(9.2078,6.81893)[tr]{$H$}
\FAProp(13.,14.)(13.,6.)(0.,){/ScalarDash}{-1}
\FALabel(14.2741,10.)[l]{$H$}
\FAVert(6.5,10.){0}
\FAVert(13.,14.){0}
\FAVert(13.,6.){0}

\FADiagram{}
\FAProp(0.,10.)(6.5,10.)(0.,){/ScalarDash}{0}
\FALabel(3.25,9.18)[t]{$h$}
\FAProp(20,15.)(13.,14.)(0.,){/Sine}{0}
\FALabel(16.2808,15.5544)[b]{$\gamma$}
\FAProp(20,5.)(13.,6.)(0.,){/Sine}{0}
\FALabel(16.2808,4.44558)[t]{$Z$}
\FAProp(6.5,10.)(13.,14.)(0.,){/ScalarDash}{-1}
\FALabel(9.2078,13.1811)[br]{$\tilde e_l^v$}
\FAProp(6.5,10.)(13.,6.)(0.,){/ScalarDash}{1}
\FALabel(9.2078,6.81893)[tr]{$\tilde e_l^w$}
\FAProp(13.,14.)(13.,6.)(0.,){/ScalarDash}{-1}
\FALabel(14.2741,10.)[l]{$\tilde e_l^v$}
\FAVert(6.5,10.){0}
\FAVert(13.,14.){0}
\FAVert(13.,6.){0}

\FADiagram{}
\FAProp(0.,10.)(6.5,10.)(0.,){/ScalarDash}{0}
\FALabel(3.25,9.18)[t]{$h$}
\FAProp(20,15.)(13.,14.)(0.,){/Sine}{0}
\FALabel(16.2808,15.5544)[b]{$\gamma$}
\FAProp(20,5.)(13.,6.)(0.,){/Sine}{0}
\FALabel(16.2808,4.44558)[t]{$Z$}
\FAProp(6.5,10.)(13.,14.)(0.,){/ScalarDash}{1}
\FALabel(9.2078,13.1811)[br]{$\tilde u_l^v$}
\FAProp(6.5,10.)(13.,6.)(0.,){/ScalarDash}{-1}
\FALabel(9.2078,6.81893)[tr]{$\tilde u_l^w$}
\FAProp(13.,14.)(13.,6.)(0.,){/ScalarDash}{1}
\FALabel(14.2741,10.)[l]{$\tilde u_l^v$}
\FAVert(6.5,10.){0}
\FAVert(13.,14.){0}
\FAVert(13.,6.){0}

\FADiagram{}
\FAProp(0.,10.)(6.5,10.)(0.,){/ScalarDash}{0}
\FALabel(3.25,9.18)[t]{$h$}
\FAProp(20,15.)(13.,14.)(0.,){/Sine}{0}
\FALabel(16.2808,15.5544)[b]{$\gamma$}
\FAProp(20,5.)(13.,6.)(0.,){/Sine}{0}
\FALabel(16.2808,4.44558)[t]{$Z$}
\FAProp(6.5,10.)(13.,14.)(0.,){/ScalarDash}{-1}
\FALabel(9.2078,13.1811)[br]{$\tilde d_l^v$}
\FAProp(6.5,10.)(13.,6.)(0.,){/ScalarDash}{1}
\FALabel(9.2078,6.81893)[tr]{$\tilde d_l^w$}
\FAProp(13.,14.)(13.,6.)(0.,){/ScalarDash}{-1}
\FALabel(14.2741,10.)[l]{$\tilde d_l^v$}
\FAVert(6.5,10.){0}
\FAVert(13.,14.){0}
\FAVert(13.,6.){0}

\FADiagram{}
\FAProp(0.,10.)(11.,10.)(0.,){/ScalarDash}{0}
\FALabel(5.5,9.18)[t]{$h$}
\FAProp(20,15.)(15.5,12.5)(0.,){/Sine}{0}
\FALabel(18.0227,12.8751)[tl]{$\gamma$}
\FAProp(20,5.)(11.,10.)(0.,){/Sine}{0}
\FALabel(15.2273,6.62506)[tr]{$Z$}
\FAProp(11.,10.)(15.5,12.5)(0.,){/Sine}{0}
\FALabel(13.5227,10.3751)[tl]{$Z$}
\FAProp(15.5,12.5)(15.5,12.5)(13.25,16.95){/ScalarDash}{-1}
\FALabel(13.0306,17.8532)[br]{$H$}
\FAVert(11.,10.){0}
\FAVert(15.5,12.5){0}

\FADiagram{}
\FAProp(0.,10.)(11.,10.)(0.,){/ScalarDash}{0}
\FALabel(5.5,9.18)[t]{$h$}
\FAProp(20,15.)(15.5,12.5)(0.,){/Sine}{0}
\FALabel(18.0227,12.8751)[tl]{$\gamma$}
\FAProp(20,5.)(11.,10.)(0.,){/Sine}{0}
\FALabel(15.2273,6.62506)[tr]{$Z$}
\FAProp(11.,10.)(15.5,12.5)(0.,){/Sine}{0}
\FALabel(13.5227,10.3751)[tl]{$Z$}
\FAProp(15.5,12.5)(15.5,12.5)(13.25,16.95){/ScalarDash}{-1}
\FALabel(13.0306,17.8532)[br]{$\tilde e_l^v$}
\FAVert(11.,10.){0}
\FAVert(15.5,12.5){0}

\FADiagram{}
\FAProp(0.,10.)(11.,10.)(0.,){/ScalarDash}{0}
\FALabel(5.5,9.18)[t]{$h$}
\FAProp(20,15.)(15.5,12.5)(0.,){/Sine}{0}
\FALabel(18.0227,12.8751)[tl]{$\gamma$}
\FAProp(20,5.)(11.,10.)(0.,){/Sine}{0}
\FALabel(15.2273,6.62506)[tr]{$Z$}
\FAProp(11.,10.)(15.5,12.5)(0.,){/Sine}{0}
\FALabel(13.5227,10.3751)[tl]{$Z$}
\FAProp(15.5,12.5)(15.5,12.5)(13.25,16.95){/ScalarDash}{-1}
\FALabel(13.0306,17.8532)[br]{$\tilde u_l^v$}
\FAVert(11.,10.){0}
\FAVert(15.5,12.5){0}

\FADiagram{}
\FAProp(0.,10.)(11.,10.)(0.,){/ScalarDash}{0}
\FALabel(5.5,9.18)[t]{$h$}
\FAProp(20,15.)(15.5,12.5)(0.,){/Sine}{0}
\FALabel(18.0227,12.8751)[tl]{$\gamma$}
\FAProp(20,5.)(11.,10.)(0.,){/Sine}{0}
\FALabel(15.2273,6.62506)[tr]{$Z$}
\FAProp(11.,10.)(15.5,12.5)(0.,){/Sine}{0}
\FALabel(13.5227,10.3751)[tl]{$Z$}
\FAProp(15.5,12.5)(15.5,12.5)(13.25,16.95){/ScalarDash}{-1}
\FALabel(13.0306,17.8532)[br]{$\tilde d_l^v$}
\FAVert(11.,10.){0}
\FAVert(15.5,12.5){0}

\FADiagram{}
\FAProp(0.,10.)(7.,10.)(0.,){/ScalarDash}{0}
\FALabel(3.5,9.18)[t]{$h$}
\FAProp(20,15.)(13.,10.)(0.,){/Sine}{0}
\FALabel(16.0791,13.2813)[br]{$\gamma$}
\FAProp(20,5.)(13.,10.)(0.,){/Sine}{0}
\FALabel(16.9209,8.28129)[bl]{$Z$}
\FAProp(7.,10.)(13.,10.)(0.833,){/ScalarDash}{-1}
\FALabel(10.,6.431)[t]{$H$}
\FAProp(7.,10.)(13.,10.)(-0.833,){/ScalarDash}{1}
\FALabel(10.,13.569)[b]{$H$}
\FAVert(7.,10.){0}
\FAVert(13.,10.){0}

\FADiagram{}
\FAProp(0.,10.)(7.,10.)(0.,){/ScalarDash}{0}
\FALabel(3.5,9.18)[t]{$h$}
\FAProp(20,15.)(13.,10.)(0.,){/Sine}{0}
\FALabel(16.0791,13.2813)[br]{$\gamma$}
\FAProp(20,5.)(13.,10.)(0.,){/Sine}{0}
\FALabel(16.9209,8.28129)[bl]{$Z$}
\FAProp(7.,10.)(13.,10.)(0.833,){/ScalarDash}{-1}
\FALabel(10.,6.431)[t]{$\tilde e_l^v$}
\FAProp(7.,10.)(13.,10.)(-0.833,){/ScalarDash}{1}
\FALabel(10.,13.569)[b]{$\tilde e_l^w$}
\FAVert(7.,10.){0}
\FAVert(13.,10.){0}

\FADiagram{}
\FAProp(0.,10.)(7.,10.)(0.,){/ScalarDash}{0}
\FALabel(3.5,9.18)[t]{$h$}
\FAProp(20,15.)(13.,10.)(0.,){/Sine}{0}
\FALabel(16.0791,13.2813)[br]{$\gamma$}
\FAProp(20,5.)(13.,10.)(0.,){/Sine}{0}
\FALabel(16.9209,8.28129)[bl]{$Z$}
\FAProp(7.,10.)(13.,10.)(0.833,){/ScalarDash}{-1}
\FALabel(10.,6.431)[t]{$\tilde u_l^v$}
\FAProp(7.,10.)(13.,10.)(-0.833,){/ScalarDash}{1}
\FALabel(10.,13.569)[b]{$\tilde u_l^w$}
\FAVert(7.,10.){0}
\FAVert(13.,10.){0}

\FADiagram{}
\FAProp(0.,10.)(7.,10.)(0.,){/ScalarDash}{0}
\FALabel(3.5,9.18)[t]{$h$}
\FAProp(20,15.)(13.,10.)(0.,){/Sine}{0}
\FALabel(16.0791,13.2813)[br]{$\gamma$}
\FAProp(20,5.)(13.,10.)(0.,){/Sine}{0}
\FALabel(16.9209,8.28129)[bl]{$Z$}
\FAProp(7.,10.)(13.,10.)(0.833,){/ScalarDash}{-1}
\FALabel(10.,6.431)[t]{$\tilde d_l^v$}
\FAProp(7.,10.)(13.,10.)(-0.833,){/ScalarDash}{1}
\FALabel(10.,13.569)[b]{$\tilde d_l^w$}
\FAVert(7.,10.){0}
\FAVert(13.,10.){0}

\FADiagram{}
\FAProp(0.,10.)(11.,10.)(0.,){/ScalarDash}{0}
\FALabel(5.5,9.18)[t]{$h$}
\FAProp(20,15.)(17.3,13.5)(0.,){/Sine}{0}
\FALabel(18.3773,15.1249)[br]{$\gamma$}
\FAProp(20,5.)(11.,10.)(0.,){/Sine}{0}
\FALabel(15.2273,6.62506)[tr]{$Z$}
\FAProp(11.,10.)(13.7,11.5)(0.,){/ScalarDash}{0}
\FALabel(12.1987,11.4064)[br]{$A$}
\FAProp(17.3,13.5)(13.7,11.5)(-0.8,){/Straight}{-1}
\FALabel(16.5727,10.1851)[tl]{$e_l$}
\FAProp(17.3,13.5)(13.7,11.5)(0.8,){/Straight}{1}
\FALabel(14.4273,14.8149)[br]{$e_l$}
\FAVert(11.,10.){0}
\FAVert(17.3,13.5){0}
\FAVert(13.7,11.5){0}

\FADiagram{}
\FAProp(0.,10.)(11.,10.)(0.,){/ScalarDash}{0}
\FALabel(5.5,9.18)[t]{$h$}
\FAProp(20,15.)(17.3,13.5)(0.,){/Sine}{0}
\FALabel(18.3773,15.1249)[br]{$\gamma$}
\FAProp(20,5.)(11.,10.)(0.,){/Sine}{0}
\FALabel(15.2273,6.62506)[tr]{$Z$}
\FAProp(11.,10.)(13.7,11.5)(0.,){/ScalarDash}{0}
\FALabel(12.1987,11.4064)[br]{$A^0$}
\FAProp(17.3,13.5)(13.7,11.5)(-0.8,){/Straight}{-1}
\FALabel(16.5727,10.1851)[tl]{$u_l$}
\FAProp(17.3,13.5)(13.7,11.5)(0.8,){/Straight}{1}
\FALabel(14.4273,14.8149)[br]{$u_l$}
\FAVert(11.,10.){0}
\FAVert(17.3,13.5){0}
\FAVert(13.7,11.5){0}

\FADiagram{}
\FAProp(0.,10.)(11.,10.)(0.,){/ScalarDash}{0}
\FALabel(5.5,9.18)[t]{$h$}
\FAProp(20,15.)(17.3,13.5)(0.,){/Sine}{0}
\FALabel(18.3773,15.1249)[br]{$\gamma$}
\FAProp(20,5.)(11.,10.)(0.,){/Sine}{0}
\FALabel(15.2273,6.62506)[tr]{$Z$}
\FAProp(11.,10.)(13.7,11.5)(0.,){/ScalarDash}{0}
\FALabel(12.1987,11.4064)[br]{$A$}
\FAProp(17.3,13.5)(13.7,11.5)(-0.8,){/Straight}{-1}
\FALabel(16.5727,10.1851)[tl]{$d_l$}
\FAProp(17.3,13.5)(13.7,11.5)(0.8,){/Straight}{1}
\FALabel(14.4273,14.8149)[br]{$d_l$}
\FAVert(11.,10.){0}
\FAVert(17.3,13.5){0}
\FAVert(13.7,11.5){0}

\FADiagram{}
\FAProp(0.,10.)(11.,10.)(0.,){/ScalarDash}{0}
\FALabel(5.5,9.18)[t]{$h$}
\FAProp(20,15.)(17.3,13.5)(0.,){/Sine}{0}
\FALabel(18.3773,15.1249)[br]{$\gamma$}
\FAProp(20,5.)(11.,10.)(0.,){/Sine}{0}
\FALabel(15.2273,6.62506)[tr]{$Z$}
\FAProp(11.,10.)(13.7,11.5)(0.,){/ScalarDash}{0}
\FALabel(12.1987,11.4064)[br]{$A$}
\FAProp(17.3,13.5)(13.7,11.5)(-0.8,){/Straight}{-1}
\FALabel(16.5727,10.1851)[tl]{$\tilde \chi_l$}
\FAProp(17.3,13.5)(13.7,11.5)(0.8,){/Straight}{1}
\FALabel(14.4273,14.8149)[br]{$\tilde \chi_l$}
\FAVert(11.,10.){0}
\FAVert(17.3,13.5){0}
\FAVert(13.7,11.5){0}

\FADiagram{}
\FAProp(0.,10.)(11.,10.)(0.,){/ScalarDash}{0}
\FALabel(5.5,9.18)[t]{$h$}
\FAProp(20,15.)(17.3,13.5)(0.,){/Sine}{0}
\FALabel(18.3773,15.1249)[br]{$\gamma$}
\FAProp(20,5.)(11.,10.)(0.,){/Sine}{0}
\FALabel(15.2273,6.62506)[tr]{$Z$}
\FAProp(11.,10.)(13.7,11.5)(0.,){/ScalarDash}{0}
\FALabel(12.1987,11.4064)[br]{$A$}
\FAProp(17.3,13.5)(13.7,11.5)(-0.8,){/ScalarDash}{-1}
\FALabel(16.5727,10.1851)[tl]{$\tilde e_l^v$}
\FAProp(17.3,13.5)(13.7,11.5)(0.8,){/ScalarDash}{1}
\FALabel(14.4273,14.8149)[br]{$\tilde e_l^v$}
\FAVert(11.,10.){0}
\FAVert(17.3,13.5){0}
\FAVert(13.7,11.5){0}

\FADiagram{}
\FAProp(0.,10.)(11.,10.)(0.,){/ScalarDash}{0}
\FALabel(5.5,9.18)[t]{$h$}
\FAProp(20,15.)(17.3,13.5)(0.,){/Sine}{0}
\FALabel(18.3773,15.1249)[br]{$\gamma$}
\FAProp(20,5.)(11.,10.)(0.,){/Sine}{0}
\FALabel(15.2273,6.62506)[tr]{$Z$}
\FAProp(11.,10.)(13.7,11.5)(0.,){/ScalarDash}{0}
\FALabel(12.1987,11.4064)[br]{$A$}
\FAProp(17.3,13.5)(13.7,11.5)(-0.8,){/ScalarDash}{-1}
\FALabel(16.5727,10.1851)[tl]{$\tilde u_l^v$}
\FAProp(17.3,13.5)(13.7,11.5)(0.8,){/ScalarDash}{1}
\FALabel(14.4273,14.8149)[br]{$\tilde u_l^v$}
\FAVert(11.,10.){0}
\FAVert(17.3,13.5){0}
\FAVert(13.7,11.5){0}

\FADiagram{}
\FAProp(0.,10.)(11.,10.)(0.,){/ScalarDash}{0}
\FALabel(5.5,9.18)[t]{$h$}
\FAProp(20,15.)(17.3,13.5)(0.,){/Sine}{0}
\FALabel(18.3773,15.1249)[br]{$\gamma$}
\FAProp(20,5.)(11.,10.)(0.,){/Sine}{0}
\FALabel(15.2273,6.62506)[tr]{$Z$}
\FAProp(11.,10.)(13.7,11.5)(0.,){/ScalarDash}{0}
\FALabel(12.1987,11.4064)[br]{$A$}
\FAProp(17.3,13.5)(13.7,11.5)(-0.8,){/ScalarDash}{-1}
\FALabel(16.5727,10.1851)[tl]{$\tilde d_l^v$}
\FAProp(17.3,13.5)(13.7,11.5)(0.8,){/ScalarDash}{1}
\FALabel(14.4273,14.8149)[br]{$\tilde d_l^v$}
\FAVert(11.,10.){0}
\FAVert(17.3,13.5){0}
\FAVert(13.7,11.5){0}

\FADiagram{}
\FAProp(0.,10.)(11.,10.)(0.,){/ScalarDash}{0}
\FALabel(5.5,9.18)[t]{$h$}
\FAProp(20,15.)(17.3,13.5)(0.,){/Sine}{0}
\FALabel(18.3773,15.1249)[br]{$\gamma$}
\FAProp(20,5.)(11.,10.)(0.,){/Sine}{0}
\FALabel(15.2273,6.62506)[tr]{$Z$}
\FAProp(11.,10.)(13.7,11.5)(0.,){/Sine}{0}
\FALabel(12.0773,11.6249)[br]{$Z$}
\FAProp(17.3,13.5)(13.7,11.5)(-0.8,){/Straight}{-1}
\FALabel(16.5727,10.1851)[tl]{$\tilde \chi_l$}
\FAProp(17.3,13.5)(13.7,11.5)(0.8,){/Straight}{1}
\FALabel(14.4273,14.8149)[br]{$\tilde \chi_l$}
\FAVert(11.,10.){0}
\FAVert(17.3,13.5){0}
\FAVert(13.7,11.5){0}

\FADiagram{}
\FAProp(0.,10.)(11.,10.)(0.,){/ScalarDash}{0}
\FALabel(5.5,9.18)[t]{$h$}
\FAProp(20,15.)(17.3,13.5)(0.,){/Sine}{0}
\FALabel(18.3773,15.1249)[br]{$\gamma$}
\FAProp(20,5.)(11.,10.)(0.,){/Sine}{0}
\FALabel(15.2273,6.62506)[tr]{$Z$}
\FAProp(11.,10.)(13.7,11.5)(0.,){/Sine}{0}
\FALabel(12.0773,11.6249)[br]{$Z$}
\FAProp(17.3,13.5)(13.7,11.5)(-0.8,){/ScalarDash}{-1}
\FALabel(16.5727,10.1851)[tl]{$H$}
\FAProp(17.3,13.5)(13.7,11.5)(0.8,){/ScalarDash}{1}
\FALabel(14.4273,14.8149)[br]{$H$}
\FAVert(11.,10.){0}
\FAVert(17.3,13.5){0}
\FAVert(13.7,11.5){0}

\FADiagram{}
\FAProp(0.,10.)(11.,10.)(0.,){/ScalarDash}{0}
\FALabel(5.5,9.18)[t]{$h$}
\FAProp(20,15.)(17.3,13.5)(0.,){/Sine}{0}
\FALabel(18.3773,15.1249)[br]{$\gamma$}
\FAProp(20,5.)(11.,10.)(0.,){/Sine}{0}
\FALabel(15.2273,6.62506)[tr]{$Z$}
\FAProp(11.,10.)(13.7,11.5)(0.,){/Sine}{0}
\FALabel(12.0773,11.6249)[br]{$Z$}
\FAProp(17.3,13.5)(13.7,11.5)(-0.8,){/ScalarDash}{-1}
\FALabel(16.5727,10.1851)[tl]{$\tilde e_l^v$}
\FAProp(17.3,13.5)(13.7,11.5)(0.8,){/ScalarDash}{1}
\FALabel(14.4273,14.8149)[br]{$\tilde e_l^v$}
\FAVert(11.,10.){0}
\FAVert(17.3,13.5){0}
\FAVert(13.7,11.5){0}

\FADiagram{}
\FAProp(0.,10.)(11.,10.)(0.,){/ScalarDash}{0}
\FALabel(5.5,9.18)[t]{$h$}
\FAProp(20,15.)(17.3,13.5)(0.,){/Sine}{0}
\FALabel(18.3773,15.1249)[br]{$\gamma$}
\FAProp(20,5.)(11.,10.)(0.,){/Sine}{0}
\FALabel(15.2273,6.62506)[tr]{$Z$}
\FAProp(11.,10.)(13.7,11.5)(0.,){/Sine}{0}
\FALabel(12.0773,11.6249)[br]{$Z$}
\FAProp(17.3,13.5)(13.7,11.5)(-0.8,){/ScalarDash}{-1}
\FALabel(16.5727,10.1851)[tl]{$\tilde u_l^v$}
\FAProp(17.3,13.5)(13.7,11.5)(0.8,){/ScalarDash}{1}
\FALabel(14.4273,14.8149)[br]{$\tilde u_l^v$}
\FAVert(11.,10.){0}
\FAVert(17.3,13.5){0}
\FAVert(13.7,11.5){0}

\FADiagram{}
\FAProp(0.,10.)(11.,10.)(0.,){/ScalarDash}{0}
\FALabel(5.5,9.18)[t]{$h$}
\FAProp(20,15.)(17.3,13.5)(0.,){/Sine}{0}
\FALabel(18.3773,15.1249)[br]{$\gamma$}
\FAProp(20,5.)(11.,10.)(0.,){/Sine}{0}
\FALabel(15.2273,6.62506)[tr]{$Z$}
\FAProp(11.,10.)(13.7,11.5)(0.,){/Sine}{0}
\FALabel(12.0773,11.6249)[br]{$Z$}
\FAProp(17.3,13.5)(13.7,11.5)(-0.8,){/ScalarDash}{-1}
\FALabel(16.5727,10.1851)[tl]{$\tilde d_l^v$}
\FAProp(17.3,13.5)(13.7,11.5)(0.8,){/ScalarDash}{1}
\FALabel(14.4273,14.8149)[br]{$\tilde d_l^v$}
\FAVert(11.,10.){0}
\FAVert(17.3,13.5){0}
\FAVert(13.7,11.5){0}
\end{feynartspicture}
\end{center}
\vspace{-1.5in}
\caption{Representative Feynman diagrams for the $h \rightarrow Z \gamma$ decay at the one-loop level in the MSSM. Here, $\tilde{u}_{l}$ and $\tilde{d}_{l}$ denote mass eigenstates of up-type and down-type scalar quarks, $\tilde{e}_{l}$ represents mass eigenstates of charged scalar leptons, $\tilde{\chi}_{l}$ indicates charginos, and $H$, $A$ are additional MSSM Higgs bosons.}
\label{FeynDiagHZga}
\end{figure}

In our initial step, we computed the analytical results for $(h \rightarrow Z \gamma)$ within the MSSM using the pre-existing model file for MSSM. The relevant Feynman diagrams involved in our calculations are shown in \reffi{FeynDiagHZga}. Here, $\tilde{u}_{l}$ and $\tilde{d}_{l}$ denote the mass eigenstates of up-type and down-type scalar quarks respectively, $\tilde{e}_{l}$ represents the mass eigenstates of charged scalar leptons, $\tilde{\chi}_{l}$ indicates charginos, and $H$, $A$ refer to additional MSSM Higgs bosons. Contributions from these Feynman diagrams, particularly those involving $\tilde{\chi}_{l}$, $H$, and $A$, have the potential to significantly enhance $\Gamma(h \rightarrow Z \gamma)$. The analytical results were further simplified using \fc, and these simplified expressions were subsequently employed for the numerical evaluation of $\Gamma(h \rightarrow Z \gamma)$.

In the case of NHSSM, the model file isn't available in \fa. We generated the NHSSM model file using the Mathematica package {\tt SARAH} and prepared the necessary driver files for \fc. Once the model file was obtained, we followed the previously mentioned steps to compute $\Gamma(h \rightarrow Z \gamma)$ within the NHSSM framework.

Unlike the MSSM, NHSSM does not introduce additional Feynman diagrams; rather, it shares the same diagrams as the MSSM. However, NHSSM includes additional couplings that can contribute to $\Gamma(h \rightarrow Z \gamma)$.  

\section{Numerical Results}
\label{sec:NResults}

\subsection{Input parameters}
\label{sec:input-para}

For our numerical analysis, we have chosen three scenarios labeled \( M_{h}^{125} \), \( M_{h}^{125}(\tilde{\tau}) \), and \( M_{h}^{125}(\tilde{\chi}) \), as originally presented in \citere{Bahl:2018zmf}. These scenarios are defined using electroweak-scale parameters, selected to emphasize different aspects of Higgs boson phenomenology within the MSSM.
It is important to note that these scenarios, over a wide range of their parameter space, are consistent with experimental results from the LHC concerning the properties of the Higgs boson, as well as with constraints on the masses and couplings of new particles. Each scenario includes a $\cp$-even scalar with a mass close to 125 GeV, exhibiting couplings similar to those of the SM.

 In these benchmark scenarios, the SSB \( M_{\tilde f} \) is set to \( 2 \tev \) for the first two generations, while the holomorphic trilinear SSB terms for these generations are assumed to be zero. The remaining parameters are given in~\refta{tab:input-parameters}~\cite{Bahl:2018zmf}.

\begin{table}[htbp] 
\centerline{\begin{tabular}{|c|c|c|c|}
\hline\hline
& $M_{h}^{125}$  & $M_{h}^{125}(\tilde{\tau})$ & $M_{h}^{125}%
(\tilde{\chi})$ \\\hline\hline
$m_{\tilde{Q}_{3}},m_{\tilde{U}_{3}},m_{\tilde{D}_{3}}$ & $1500$ & $1500$ &
$1500$\\
$m_{\tilde{L}_{3}},m_{\tilde{E}_{3}}$ & $2000$ & $350$ & $2000$\\
$\mu$ & $1000$ & $1000$ & $180$\\
$M_{1}$ & $1000$ & $180$ & $160$\\
$M_{2}$ & $1000$ & $300$ & $180$\\
$M_{3}$ & $2500$ & $2500$ & $2500$\\
$X_{t}$ & $2800$ & $2800$ & $2500$\\
$A_{\tau}$ & $0$ & $800$ & $0$\\
\hline\hline
\end{tabular}}
\caption{Selected scenarios in the MSSM parameter space, taken from
  \citere{Bahl:2018zmf}. All the dimensionful quantities are in $\gev$.}%
\label{tab:input-parameters}
\end{table}%

Here, the parameters \( m_{\tilde{Q}_{3}} \), \( m_{\tilde{U}_{3}} \), and \( m_{\tilde{D}_{3}} \) correspond to the masses of the third generation squark doublet, up-type squark singlet, and down-type squark singlet, respectively. Additionally, \( m_{\tilde{L}_{3}} \) and \( m_{\tilde{E}_{3}} \) represent the masses of the third generation left-handed slepton doublet and right-handed slepton singlet, respectively. The parameters \( X_t = A_t - \mu \cot \beta \) and \( X_\tau = A_\tau - \mu \cot \beta \) are the holomorphic trilinear couplings.

While defining these scenarios, certain indirect constraints, such as those related to dark matter relic density and the anomalous magnetic moment of the muon, were not taken into account, as their influence on parameters linked to Higgs phenomenology is expected to be negligible.

\subsection{$\Gamma(h\rightarrow Z \gamma)$ in MSSM}
\label{sec:WidthHZga-MSSM}

The SM prediction for $\Gamma(h \rightarrow Z \gamma)$ deviates from the experimental value by about -56\% (please see \refse{Exp-status}). Therefore, it is crucial to identify the sources that can minimize this deviation and reconcile theoretical predictions with experimental results. In this section, we will examine whether the MSSM can fulfill this task.

\begin{figure}
\centering
\includegraphics[width=6.5cm,keepaspectratio]{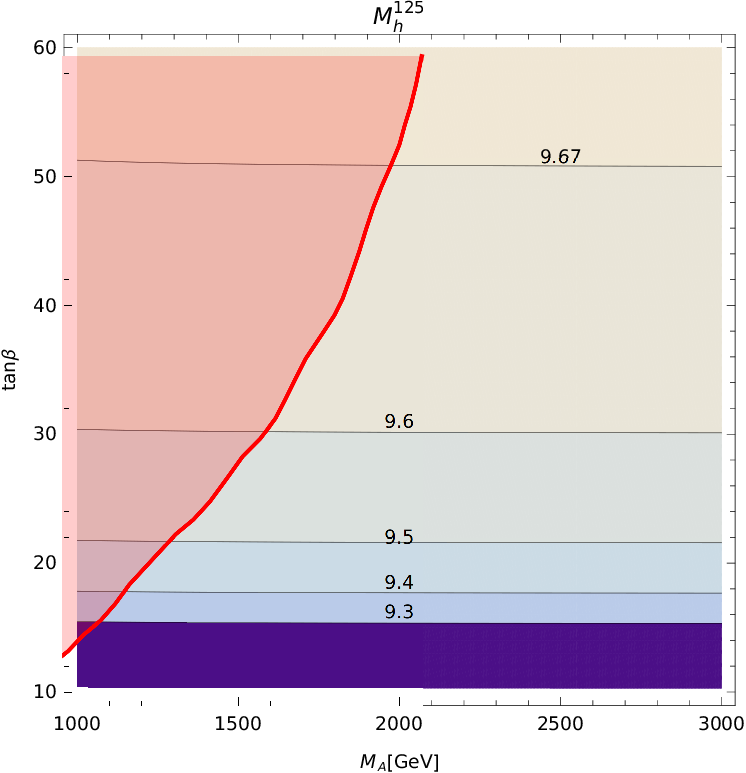}
\hspace{0.02\textwidth}
\includegraphics[width=6.5cm,keepaspectratio]{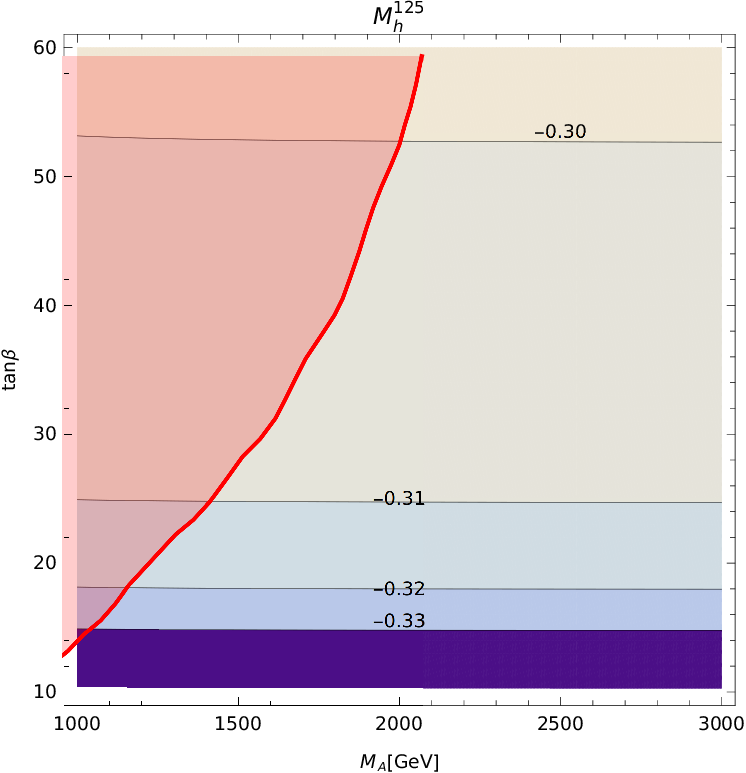}\\
\vspace{0.02\textwidth}
\includegraphics[width=6.5cm,keepaspectratio]{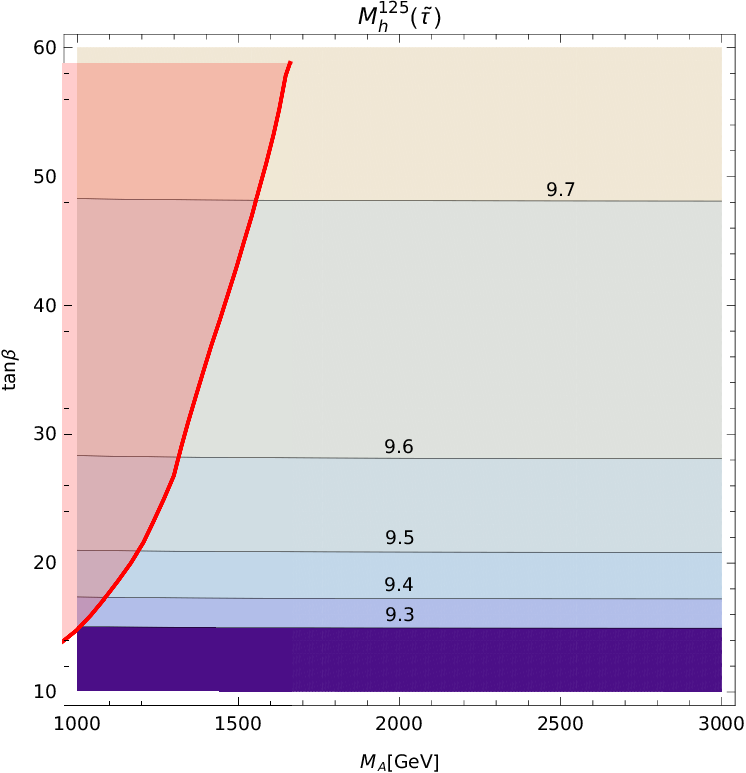}
\hspace{0.02\textwidth}
\includegraphics[width=6.5cm,keepaspectratio]{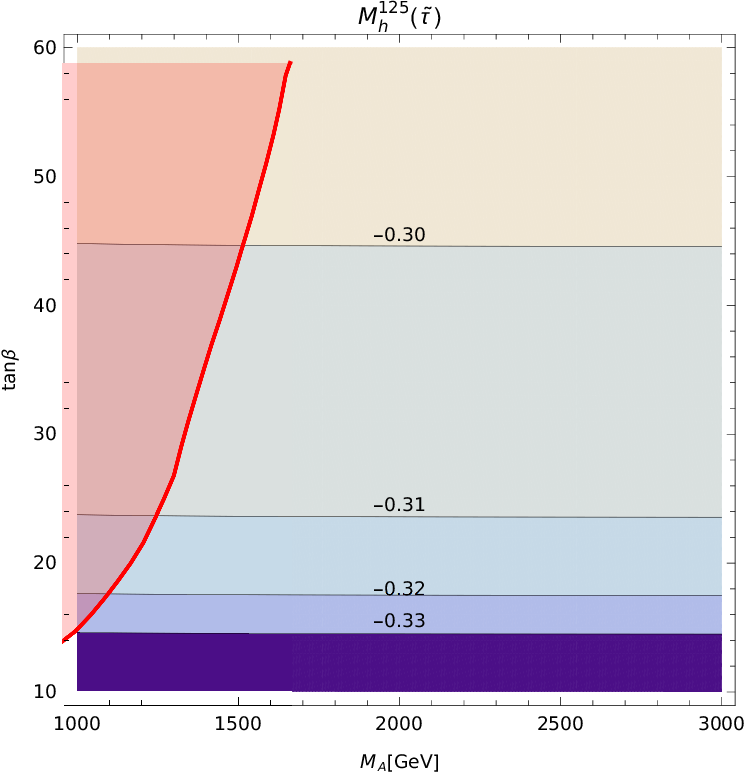}\\
\vspace{0.02\textwidth}
\includegraphics[width=6.5cm,keepaspectratio]{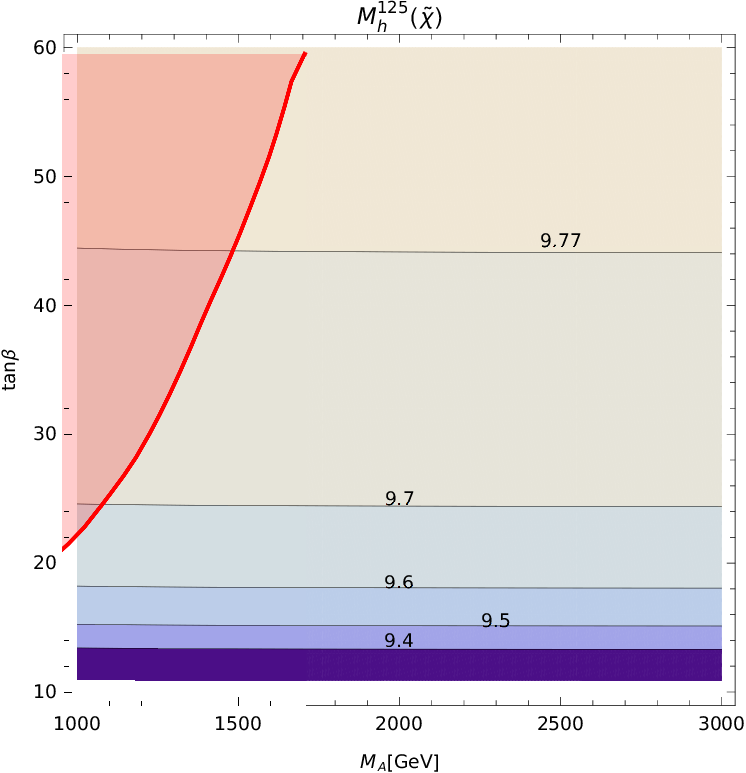}
\hspace{0.02\textwidth}
\includegraphics[width=6.5cm,keepaspectratio]{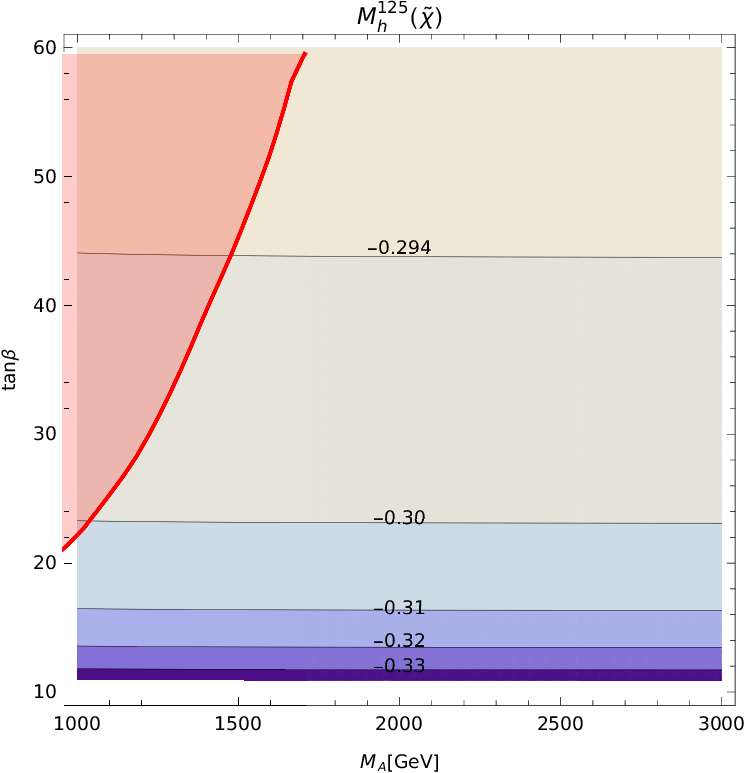}\\

\caption{$\Gamma(h \rightarrow Z \gamma)_{\rm MSSM} \times 10^{6} \gev$ (left column) and ${\rm Dev}(Z \gamma)_{\rm MSSM}$ (right column) in the $(M_A, \tb)$ plane within the $M_{h}^{125}$ (upper row), $M_{h}^{125}(\tilde{\tau})$ (middle row) and $M_{h}^{125}(\tilde{\chi})$ (lower row) scenarios.}
\label{fig:MSSM}
\end{figure}

Our results are presented in \reffi{fig:MSSM}. In the left panel, we show $\Gamma(h \rightarrow Z \gamma)_{\rm MSSM}$ multiplied by $10^{6}$, and in the right panel, we show ${\rm Dev}(Z \gamma)_{\rm MSSM}$ as defined in \refeq{Eq:Dev-MSSM}, both plotted in the $(M_A, \tan\beta)$ plane for the three selected scenarios. The plots in the upper row correspond to the $M_{h}^{125}$ scenario, the middle row depicts the $M_{h}^{125}(\tilde{\tau})$ scenario, and the bottom row shows the $M_{h}^{125}(\tilde{\chi})$ scenario. The red shaded area is excluded by direct searches at the LHC \cite{Bahl:2018zmf,ATLAS:2020zms,CMS:2022goy,ATLAS:2024lyh}.

As seen from the left panels in \reffi{fig:MSSM}, $\Gamma(h \rightarrow Z \gamma)_{\rm MSSM}$ can reach a maximum value of approximately $9.70 \times 10^{-6} \, \text{GeV}$ in the $M_{h}^{125}$ scenario. This value is slightly higher in the other two scenarios, reaching up to $9.77 \times 10^{-6} \, \text{GeV}$. Compared to the SM prediction $\Gamma(h \rightarrow Z \gamma)_{\text{SM}} = (6.1 \pm 0.4) \times 10^{-6} \, \text{GeV}$, these values are closer to the experimentally measured value. The deviation reduces to -30\% for the $M_{h}^{125}$ scenario (see upper right plot) and to approximately -29.4\% for the other two scenarios considered here (see middle and lower right plots). This deviation is well within $1\sigma$, which was about $1.7\sigma$ for the SM. If this signal is confirmed at the LHC, it could be a potential hint for the existence of the MSSM.

\subsection{$\Gamma(h\rightarrow Z \gamma)$ in NHSSM} 

In this section, we present our numerical results for the NHSSM. For each scenario, we investigated four different combinations of $M_A$ and $\tan\beta$, considering the latest experimental limits from MSSM Higgs boson searches.
~\cite{ATLAS:2020zms,CMS:2022goy,ATLAS:2024lyh}:
\begin{align*}
\rm P1 &:~\MA = 1500 \gev,~ \tb = 7 \\
\rm P2 &:~\MA = 2000 \gev,~ \tb = 15 \\
\rm P3 &:~\MA = 2500 \gev,~ \tb = 30 \\
\rm P4 &:~\MA = 2500 \gev,~ \tb = 45 
\end{align*}

\begin{figure}
\centering
\includegraphics[width=7.5cm,keepaspectratio]{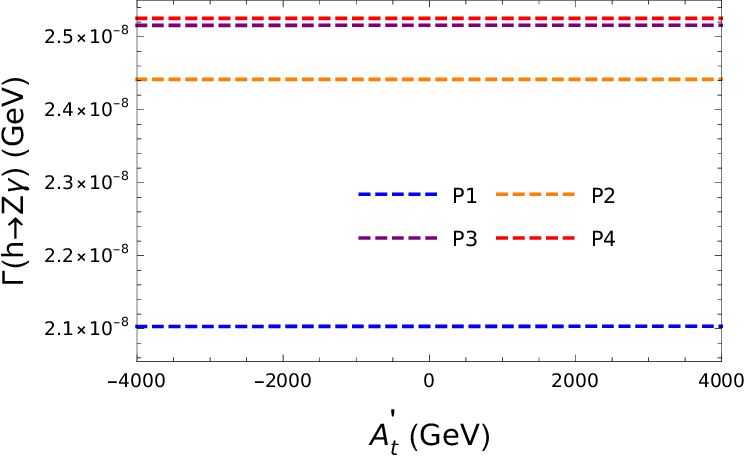}
\hspace{0.02\textwidth}
\includegraphics[width=7.5cm,keepaspectratio]{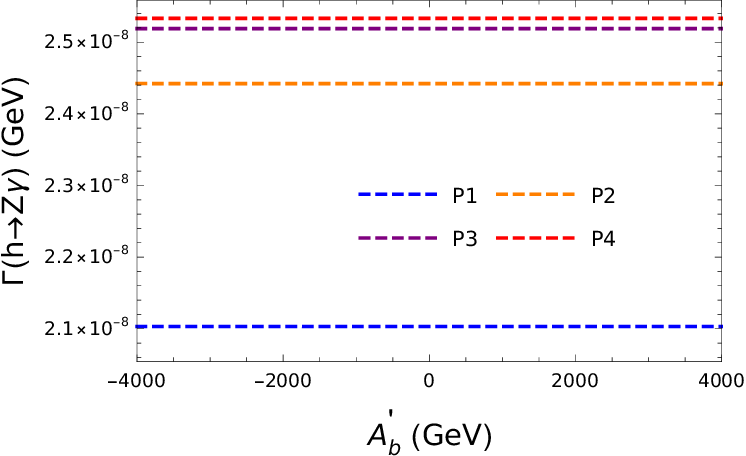}\\
\vspace{0.02\textwidth}
\includegraphics[width=7.5cm,keepaspectratio]{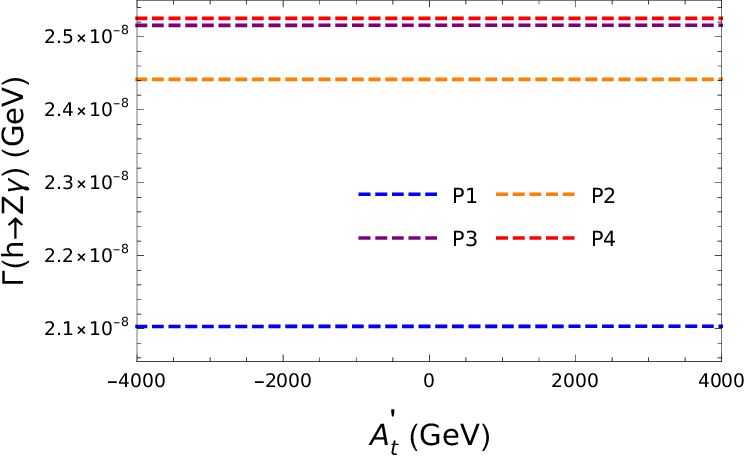}
\hspace{0.02\textwidth}
\includegraphics[width=7.5cm,keepaspectratio]{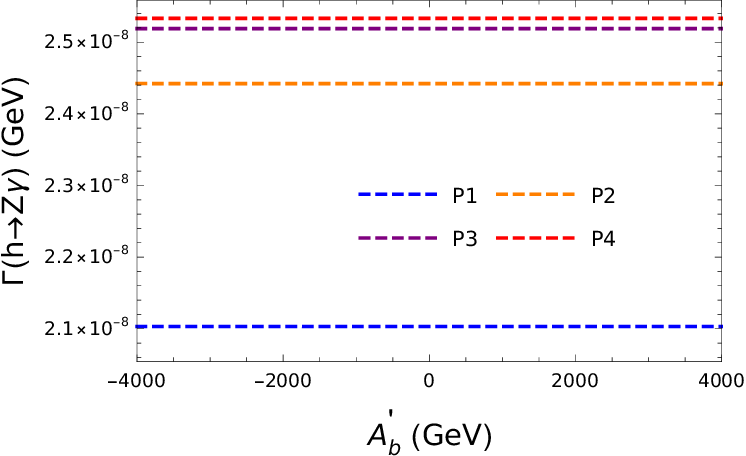}\\
\vspace{0.02\textwidth}
\includegraphics[width=7.5cm,keepaspectratio]{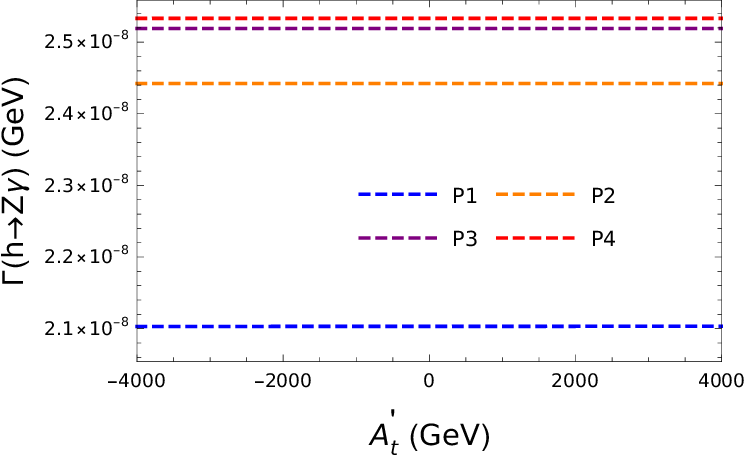}
\hspace{0.02\textwidth}
\includegraphics[width=7.5cm,keepaspectratio]{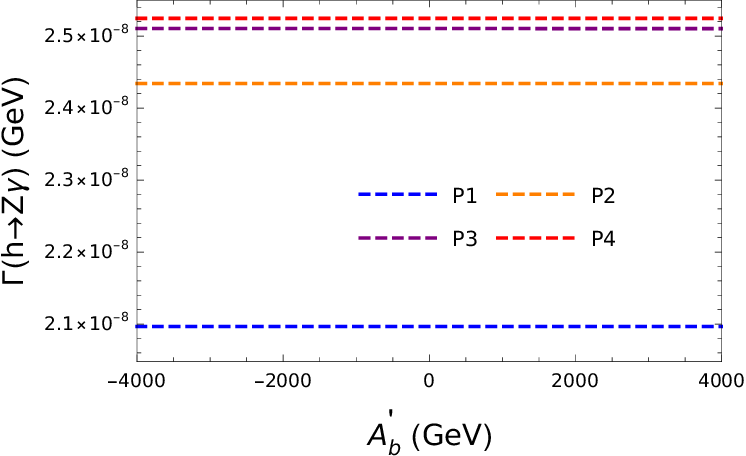}\\

\caption{$\Gamma(h \rightarrow Z \gamma)_{\rm NHSSM}$ as a function of $A_t^{\prime}$ (left column) and $A_b^{\prime}$ (right column) for $M_{h}^{125}$ (upper row), $M_{h}^{125}(\tilde{\tau})$ (middle row) and $M_{h}^{125}(\tilde{\chi})$ (lower row) scenarios.}
\label{fig:NHSSM}
\end{figure}

Nonholomorphic terms only affect the couplings of scalar quarks and scalar leptons to other particles. Therefore, for the NHSSM analysis, we selected only the Feynman diagrams that include scalar quarks and scalar leptons in the loop, as these are the only diagrams impacted by these terms. Our results are shown in \reffi{fig:NHSSM}, where we present the nonholomorphic contributions to $\Gamma(h \rightarrow Z \gamma)_{\rm NHSSM}$ as a function of $A_t^{\prime}$ (left panel) and $A_b^{\prime}$ (right panel). The upper row contains plots for the $M_{h}^{125}$ scenario, while the middle and lower rows correspond to the $M_{h}^{125}(\tilde{\tau})$ and $M_{h}^{125}(\tilde{\chi})$ scenarios, respectively.

The value of $\Gamma(h \rightarrow Z \gamma)_{\rm NHSSM}$ at $A_t^{\prime} = 0$ and $A_b^{\prime} = 0$ corresponds to the scalar quark contribution in the MSSM. As can be seen from the \reffi{fig:NHSSM}, the scalar quark contributions to $\Gamma(h \rightarrow Z \gamma)$ are two orders of magnitude smaller compared to the total contributions in the MSSM, reaching only $\order{10^{-8}}$. The largest contributions are obtained for the point P4, and the contributions decrease with decreasing values of $\tan\beta$, confirming the same trend observed in the previous figure. 

Furthermore, the effects of the NHSSM couplings are negligible for these contributions, as evidenced by the flat curves in all the plots. This indicates that the inclusion of nonholomorphic terms does not significantly alter the scalar quark contributions to $\Gamma(h \rightarrow Z \gamma)$ in the NHSSM.

\clearpage
\section{Conclusions}
\label{sec:conclusions}
In this work, we have calculated $\Gamma(h \rightarrow Z \gamma)$ in the Minimal Supersymmetric Standard Model (MSSM) and its nonholomorphic extension, the NHSSM. The calculations were performed using the \fa/\fc~ framework. We utilized the existing \fa~ model file for the MSSM and developed a new model file, along with the necessary driver files, using the Mathematica package {\tt SARAH}.

For the numerical analysis, we selected three benchmark scenarios namely the $M_{h}^{125}$, $M_{h}^{125}(\tilde{\tau})$ and $M_{h}^{125}(\tilde{\chi})$ and examined MSSM contributions to $\Gamma(h \rightarrow Z \gamma)$ in the $(M_A, \tan\beta)$ plane. The MSSM contributions can be quite significant, reaching $9.70 \times 10^{-6} \, \text{GeV}$ in the $M_{h}^{125}$ scenario and up to $9.77 \times 10^{-6} \, \text{GeV}$ in the $M_{h}^{125}(\tilde{\tau})$ and $M_{h}^{125}(\tilde{\chi})$ scenarios. These results are closer to the experimentally predicted value, reducing the deviation to about -30\%, compared to -56\% for the Standard Model (SM). The MSSM predictions fall within the $1\sigma$ standard deviation of the experimentally measured value.

We also investigated the NHSSM contributions to $\Gamma(h \rightarrow Z \gamma)$ and presented our results for four selected combinations of $(M_A, \tan\beta)$: $(1500 \, \text{GeV}, 7)$, $(2000 \, \text{GeV}, 15)$, $(2500 \, \text{GeV}, 30)$, and $(2500 \, \text{GeV}, 45)$, which are permitted by current Higgs boson LHC searches. The NHSSM contributions were found to be negligible and do not significantly alter the predictions for $\Gamma(h \rightarrow Z \gamma)$. Despite their insignificance, including NHSSM contributions in this analysis helps reduce uncertainty in theoretical predictions.

The indications of experimental excess over the SM predictions in the $h \rightarrow Z \gamma$ decay channel, if confirmed, would serve as a strong signal for the existence of new physics, potentially including the MSSM, as demonstrated in this study.

\subsection*{Acknowledgments}

 We extend our gratitude to S. Heinemeyer for the valuable discussions that contributed to the preparation of this manuscript.


\newpage
\pagebreak

\end{document}